\documentclass[a4paper,11pt]{article} 
\pdfoutput=1
\usepackage{graphicx,rotating,hyperref,slashed,amsmath,xcolor,amssymb,amsfonts,colortbl,cite, subfigure,float} 
\makeatletter 
\usepackage{graphics} 
\usepackage{graphicx} 
\graphicspath{{plots/}}
 
\hypersetup{colorlinks,bookmarksopen,bookmarksnumbered,
linkcolor=blus,pdfstartview=FitH,urlcolor=rossos,citecolor=verde}
\numberwithin{equation}{section}
\usepackage{amsmath, tcolorbox}
\hypersetup{%
    ,urlcolor=rossos
    ,citecolor=rossos
    ,linkcolor=rossos
    }

\AtBeginDocument{
  \hypersetup{
    urlcolor=verdes,
    citecolor=verdes,
    linkcolor=verdes,
  }%
}

\allowdisplaybreaks

\def\lsim{\mathrel{\rlap{\lower3pt\hbox{\hskip0pt$\sim$}}
   \raise1pt\hbox{$<$}}}         
\def\gsim{\mathrel{\rlap{\lower4pt\hbox{\hskip1pt$\sim$}}
   \raise1pt\hbox{$>$}}}         

 \newcommand{\sfootnote}[1]{} 
\definecolor{bluc}{cmyk}{1,1,0,0.1}
\definecolor{rossoCP3}{cmyk}{0,.88,.77,.40}
\definecolor{rosso}{cmyk}{0,1,1,0.4}
\definecolor{rossos}{cmyk}{0,1,1,0.55}
\definecolor{rossoc}{cmyk}{0,1,1,0.2}
\definecolor{verdes}{cmyk}{0.92,0,0.59,0.4}

\hypersetup{colorlinks, bookmarksopen, bookmarksnumbered,
citecolor=verdes, linkcolor=bluc, pdfstartview=FitH, urlcolor=rossos}

\newcommand{\mio}[1]{}

\definecolor{Gray}{gray}{0.95}

\usepackage{multicol}
\usepackage{color}
\definecolor{rosso}{cmyk}{0,1,1,0.4}
\definecolor{rossos}{cmyk}{0,1,1,0.55}
\definecolor{rossoc}{cmyk}{0,1,1,0.2}
\definecolor{blu}{cmyk}{1,1,0,0.3}
\definecolor{blus}{cmyk}{1,1,0,0.6}
\definecolor{bluc}{cmyk}{1,1,0,0.1}
\definecolor{verde}{cmyk}{0.92,0,0.59,0.25}
\definecolor{verdec}{cmyk}{0.92,0,0.59,0.15}
\definecolor{verdes}{cmyk}{0.92,0,0.59,0.4}

\def\circa#1{\,\raise.3ex\hbox{$#1$\kern-.75em\lower1ex\hbox{$\sim$}}\,}

\newcommand{\beq}{\begin{equation}}
\newcommand{\eeq}{\end{equation}}

\newcommand{\bea}{\begin{eqnarray}}
\newcommand{\eea}{\end{eqnarray}}
\newcommand{\be}{\begin{equation}}
\newcommand{\ee}{\end{equation}}

\newfam\rsfsfam
\def\mathscr#1{{\fam\rsfsfam\relax#1}}

\def\circa#1{\,\raise.3ex\hbox{$#1$\kern-.75em\lower1ex\hbox{$\sim$}}\,}
\makeatletter

\def\hhref#1{\href{http://arxiv.org/abs/#1}{arXiv:#1}} 

\newcommand{\doi}[1]{\href{http://dx.doi.org/#1}{[doi]}}

\def\hhref#1{\href{http://arxiv.org/abs/#1}{arXiv:#1}} 
 
\def\art{\@ifnextchar[{\eart}{\oart}}
\def\eart[#1]#2#3#4#5#6{{\rm #2}, {\em #3 \bf #4} {\rm (#6) #5} ({\em #1})}

\def\article{\@ifnextchar[{\earticle}{\oarticle}}
\def\oarticle#1#2#3#4#5#6{{\rm #1}, {\em ``#6''}, {\rm #2 #3 (#5) #4}}
\def\earticle[#1]#2#3#4#5#6#7{{\rm #2}, {\em ``#7''}, {\rm #3 #4 (#6) #5}  [\hhref{#1}]}
\def\hepart[#1]#2{{\rm #2, \em#1}}
\def\heparticle[#1]#2#3{#2, {\em ``#3''} [\hhref{#1}]}

\newcounter{alphaequation}[equation]
\def\thealphaequation{\theequation\hbox to
0.6em{\hfil\alph{alphaequation}\hfil}}
\def\eqnsystem#1{
\def\@eqnnum{{\rm (\thealphaequation)}}
\def\@@eqncr{\let\@tempa\relax \ifcase\@eqcnt \def\@tempa{& & &} \or
  \def\@tempa{& &}\or \def\@tempa{&}\fi\@tempa
  \if@eqnsw\@eqnnum\refstepcounter{alphaequation}\fi
\global\@eqnswtrue\global\@eqcnt=0\cr}
\refstepcounter{equation} \let\@currentlabel\theequation \def\@tempb{#1}
\ifx\@tempb\empty\else\label{#1}\fi
\refstepcounter{alphaequation}
\let\@currentlabel\thealphaequation
\global\@eqnswtrue\global\@eqcnt=0 \tabskip\@centering\let\\=\@eqncr
$$\halign to \displaywidth\bgroup \@eqnsel\hskip\@centering
$\displaystyle\tabskip\z@{##}$&\global\@eqcnt\@ne
\hskip2\arraycolsep\hfil${##}$\hfil& \global\@eqcnt\tw@\hskip2\arraycolsep
$\displaystyle\tabskip\z@{##}$\hfil
\tabskip\@centering&\llap{##}\tabskip\z@\cr}
\def\endeqnsystem{\@@eqncr\egroup$$\global\@ignoretrue} \makeatother

\definecolor{fiorentina}{rgb}{.5,0,.5}

\begin{document}

\vspace{1truecm}
\begin{center}
\boldmath

 \huge{ \bf
{New gravitational wave probe \\ of vector dark matter}}
\unboldmath
\end{center}
\unboldmath

\vspace{-0.2cm}

\begin{center}
{\fontsize{12}{30}\selectfont  
Alisha Marriott-Best$^{a}$ \footnote{\texttt{2347066.at.swansea.ac.uk}}, Marco Peloso$^{b}$ \footnote{\texttt{marco.peloso.at.infn.it}},
 Gianmassimo Tasinato$^{a, c}$ \footnote{\texttt{g.tasinato2208.at.gmail.com}}
} 
\end{center}

\begin{center}

\vskip 8pt
\textsl{$^{a}$ Physics Department, Swansea University, SA2 8PP, United Kingdom}\\
\textsl{$^{b}$
Dipartimento di Fisica e Astronomia, Universit\`a degli Studi di Padova,\\
 INFN, Sezione di Padova, via Marzolo 8, I-35131 Padova, Italy}
\\
\textsl{$^{c}$ Dipartimento di Fisica e Astronomia, Universit\`a di Bologna,\\
 INFN, Sezione di Bologna, I.S. FLAG, viale B. Pichat 6/2, 40127 Bologna,   Italy}
\vskip 7pt

\end{center}
\smallskip
\begin{center}
\begin{abstract}
\noindent 
The longitudinal components of massive vector fields generated during inflation constitute a well-motivated dark matter candidate, with interesting  phenomenological implications. During the epoch of radiation domination following inflation, their spectrum exhibits a peak at small scales, whose  amplitude and position are governed by the parameters of the dark matter model.
We calculate the stochastic gravitational wave spectrum induced at second order in fluctuations by such a longitudinal vector peak. We   demonstrate that the amplitude of the  gravitational wave spectrum can, in principle, reach  significant values at nano-Hertz frequencies or lower. This result suggests a novel gravitational wave probe to test inflationary vector dark matter scenarios, independent from assumptions on the coupling of  dark vectors to the Standard Model.
Additionally, we derive new analytical formulas for the longitudinal vector transfer functions during radiation domination, offering a valuable tool for characterising the convolution integrals that govern the properties of the induced gravitational waves.
\end{abstract}
\end{center}

\section{Introduction}

Determining the nature of dark matter (DM) is one of the most pressing open questions
in physics. If DM is a particle, it should
couple very feebly to the Standard Model (SM) of particle physics, as only its gravitational interactions with normal matter have been observed so far \cite{Bertone:2004pz,Feng:2010gw}.  It is possible for DM particles to have no gauge interactions at all with the SM; at most, they may have kinetic mixings or interactions with gauge singlets. In this case, we can reveal its presence through so-called DM portals, see e.g. \cite{Fabbrichesi:2020wbt,Antypas:2022asj}  for reviews.
In this work we discuss a new gravitational wave probe of certain well-motivated models of particle DM. 

We consider the framework developed in \cite{Graham:2015rva} where DM consists of the longitudinal components of a massive dark photon and is produced via gravitational effects during the inflationary epoch. Gravitational production of scalar fields in the context of inflationary scenarios has previously been investigated (see \cite{Ford:1986sy,Yajnik:1990un,Chung:1998zb}). Ref.~\cite{Graham:2015rva} extended these analyses to the spin-1 case, establishing a connection to DM. Specifically, it highlighted the role of the longitudinal component induced by the mass term, whose Lagrangian in a Friedmann–Lemaître–Robertson–Walker (FLRW) background was first provided and analysed in \cite{Himmetoglu:2008hx}. The scenario proposed in \cite{Graham:2015rva} has been further examined and expanded in various theoretical studies, such as \cite{Bastero-Gil:2018uel,Ema:2019yrd,Nakai:2020cfw,Ahmed:2020fhc,Kolb:2020fwh,Salehian:2020asa,Moroi:2020has,Arvanitaki:2021qlj,Sato:2022jya,Barman:2021qds,Redi:2022zkt,Ozsoy:2023gnl}. Its potential phenomenological implications have been explored in works such as \cite{An:2014twa,McDermott:2019lch,Lee:2020wfn,Amin:2022pzv,Siemonsen:2022ivj,East:2022ppo,Pierce:2018xmy,Nomura:2019cvc,PPTA:2022eul,Unal:2022ooa,Yu:2023iog}.

When evaluated during radiation domination, the resulting momentum-dependent spectrum of longitudinal vector modes exhibits a peak at small scales relative to the CMB. The amplitude and position of this peak depend on the vector mass and other defining model parameters (see Figure \ref{fig:transfer}). We compute the second-order gravitational waves (GWs) induced by this peak in isocurvature longitudinal modes. The resulting stochastic gravitational wave background (SGWB) serves as a distinctive signature, offering a potential probe of a mechanism where DM interacts with the SM solely through gravity, making it undetectable except via its gravitational effects. If observed, the predicted momentum dependence of this GW signal would serve as a smoking gun for this scenario.

Induced GWs generated by enhanced adiabatic modes during inflation have been extensively studied~\cite{Matarrese:1993zf,Matarrese:1997ay,Nakamura:2004rm,Saito:2009jt,Espinosa:2018eve,Inomata:2019yww}. For a comprehensive review, see~\cite{Domenech:2021ztg}, as well as
\cite{LISACosmologyWorkingGroup:2025vdz}, for a recent account. This possibility is particularly well-motivated in models that produce primordial black holes (PBHs), as highlighted in the survey~\cite{Ozsoy:2023ryl}. In particular, the viable sub-lunar mass range where PBHs could constitute the entirety of DM corresponds to a GW signal that would be distinctly observable by LISA~\cite{Bartolo:2018evs,LISACosmologyWorkingGroup:2023njw}. Second-order GWs arising from isocurvature fluctuations have also been investigated, and discussed in~\cite{Domenech:2023jve,Domenech:2021and,Passaglia:2021jla}.

In this work, we explore the novel scenario where the sourcing perturbations are fluctuations in the longitudinal mode of a massive vector field. Specifically, in Section~\ref{sec_rev}, we review the framework of \cite{Graham:2015rva} and examine potential generalisations, introducing new analytical approximations for the transfer functions of the longitudinal vector mode during radiation domination. In Section~\ref{sec_SGWB}, we compute the spectrum of induced GWs in this regime, demonstrating that it exhibits a broad peak at frequencies determined by the position of the longitudinal vector spectrum peak. Finally, in Section~\ref{sec_pheno}, we discuss the phenomenological implications for GW physics.
 
Regardless of the actual realisation or extension of the  ideas introduced in \cite{Graham:2015rva}, if the longitudinal vector modes constitute all of DM, we find a well specific relation between the amplitude of the GW spectrum and the position
 of its peak, scaling as
 \be
 \nonumber
 \frac{\Omega_{\rm GW}^{\rm peak}}{10^{-10}}\,\simeq\,\left( \frac{f_{\rm peak}}{10^{-10}\,{\rm Hz}}\right)^{-2}\,.
 \ee
 We discuss how this `consistency relation' indicates that the induced GW spectrum can be measured if we have sufficient
 sensitivity to GW at nano-Hz scales, or below. We conclude with an outlook in Section \ref{sec_con}.  
 
\section{Longitudinal vector dark matter from inflation}
\label{sec_rev}

We begin with a review of the ideas developed in \cite{Graham:2015rva}, which propose a scenario for DM in terms of longitudinal vector modes produced during inflation.
We emphasise aspects which will be relevant for our later discussion of  the SGWB induced by vector longitudinal fluctuations. We  present 
new results concerning analytical fits for  transfer functions of the longitudinal vector components during radiation domination.

\smallskip

We consider a system based on a massive vector field minimally coupled with gravity. The action is:
\be
\label{eq_dia}
S\,=\,\int \sqrt{-g}
\left[
\frac{R}{2}-\frac14\,F_{\mu\nu}\,F^{\mu\nu}-\frac{M^2}{2} A_\mu A^\mu\right]\,,
\ee
where we have set $M_{\rm Pl}=1$ unless otherwise stated; $F_{\mu \nu} = \partial_\mu A_\nu - \partial_\nu A_\mu$ is the vector field strength, and $M^2$ its mass squared, that we assume to be positive to avoid ghosts~\cite{Himmetoglu:2008hx}. We consider a FLRW geometry, with scale factor $a$ plus transverse and traceless tensor perturbations (GW):
\be
d s^2\,=\,a^2(\tau) \left[ 
- d \tau^2+\left( \delta_{ij}+h_{ij} \right) d x^i d x^j
\right]\,.
\ee
Finally, we assume that the vector field has a vanishing background value, and we parametrise its fluctuations as
\be\label{eq_vede}
A_\mu\,=\,\left( A_0, \, \partial_i \varphi + A_i^T \right)\,.
\ee
The $A_0$ component is a constrained field whose evolution (as detailed below) is governed by the dynamics of the component $\varphi$, referred to as the \textit{longitudinal scalar component}. The field $A_i^T$ is the transverse vector component, which is only gravitationally coupled to the longitudinal one, and plays no relevant role in our study~\cite{Graham:2015rva}. We note that the above line element neglects the scalar fluctuations of the metric, under the assumption that  they provide subdominant contributions to the dynamics of the longitudinal vector modes. Such hypothesis is motivated by the fact that they are not amplified by the mechanism
we are going to review next. We note that we are therefore assuming that the longitudinal vector fluctuations are non-adiabatic. 

It is convenient to work in Fourier space, where the spatial part of~(\ref{eq_vede}) can be written as the sum
$\vec{A}=\vec{A}^T+ \hat k A_L$ of a transverse $\vec{A}^T$ and of a longitudinal vector component $A_L$, with 
\be
\label{rel_aL}
A_L \,=\,i \, k \, \varphi\,.
\ee
The temporal component $A_0$ appears in the action without time derivatives. In Fourier space, its equation of motion is algebraic in this variable, allowing $A_0$ to be expressed in terms of $\varphi$ (or equivalently, $A_L$) without introducing a new dynamical degree of freedom.
\be
\label{rel_aze}
A_{0 \, {\bf  k}   }\,=\,\frac{k^2}{k^2+M^2 a^2}\, \varphi'_{\bf  k} 
\,=\,\frac{-i\,k}{k^2+M^2 a^2}\, A'_{L\,\bf  k} \,,
\ee
where prime denotes differentiation with respect to conformal time. We  substitute this expression
into the quadratic action for the longitudinal scalar fluctuations in Fourier space and obtain the action for the longitudinal degree of freedom on a FLRW background~\cite{Himmetoglu:2008hx}
\be
\label{eq_dia2}
S\,=\,\int d \tau d^3 k\,\frac{k^2 a^2(\tau)}{2} \left[\frac{M^2}{k^2+M^2 a^2(\tau)}
\varphi_{\bf k}'\varphi_{\bf -k}' -M^2 \varphi_{\bf k}\varphi_{\bf -k}
\right]\,.
\ee
The longitudinal scalar variable $\varphi$ has a non-canonical kinetic structure  -- a consequence of the procedure of integrating out the auxiliary field $A_0$. The corresponding equation of motion reads
\be
\label{eveqf}
\varphi_{\bf k}''+ \frac{2 k^2\,a H}{k^2+a^2 M^2} \varphi_{\bf k}'+(k^2+M^2 a^2)\,\varphi_{\bf k}\,=\,0\,.
\ee
Building on \cite{Graham:2015rva,Ema:2019yrd,Ahmed:2020fhc,Kolb:2020fwh}, we solve this equation during inflation, and during the following radiation dominated era. For this study, it is convenient to introduce the canonically normalised mode
\begin{equation}
\pi_{\bf k} \,\equiv\,
\frac{k M a}{\sqrt{k^2+ a^2 \, M^2}} \varphi_{\bf k}\,.
\label{pi-can}
\end{equation}
Substituting this relation, the action (\ref{eq_dia2}) takes the form
\begin{equation}
S = \frac{1}{2} \int d \tau d^3 k \left[ \pi'_{\bf k} \pi'_{\bf k} - \left( k^2 + a^2 M^2 + \frac{3 k^2 M^2 a^{'2}}{\left( k^2 + a^2 M^2 \right)^2} - \frac{k^2}{k^2+a^2 M^2} \frac{a''}{a} \right) \pi_{\bf k} \pi_{\bf -k} \right] \;,
\label{ac-can}
\end{equation}
where a boundary term has been subtracted to eliminate terms with a single time derivative acting on the mode.

\subsection{Evolution
during inflation}

We assume that standard slow-roll inflation is driven by an additional field independent of the vector sector under consideration. Details of this sector are not required for our study — neither at the level of the background evolution nor at the level of fluctuations. Moreover, we can work to leading order in the slow-roll approximation, assuming a de Sitter expansion during inflation with a constant Hubble parameter $H_I$. We also assume the inequality
\be
\frac{M}{H_I}\ll 1\,,
\ee
relating the vector mass entering the action \eqref{eq_dia}, and the inflationary Hubble parameter. 

As demonstrated in~\cite{Graham:2015rva}, for very small $M$ the late-time spectrum of longitudinal perturbations during radiation domination exhibits a peak at the scales $k = a_* \, M$, where $a_*$ is the scale factor when the Hubble rate (well after inflation) is equal to $M$. Consequently, for all the scales of our interest, the condition $a M \ll k$ holds throughout inflation, both in the sub-horizon ($k \gg a H_I$) and super-horizon ($k \ll a H_I$) regimes. Namely, we consider the inequalities
\be
{\rm{subhorizon:}} \hskip1cm
\frac{k}{a \, H_I}\gg 1\gg \frac{M}{H_I} \;, 
\ee
and
\be
{\rm{superhorizon:}} \hskip1cm
\frac{M}{H_I}
\ll \frac{k}{a \, H_I} \ll 1\,,
\ee
all throughout inflation. Meaning the redefinition (\ref{pi-can}) can be approximated as 
\be
\label{relpphi}
\pi_{\bf k}\, \simeq \, a \, M \,\varphi_{\bf k}\,.
\ee
and the canonical action~(\ref{ac-can}) is well approximated by setting 
$M=0$ in the equation. In doing so the standard action of a free massive field in de Sitter is obtained, which then provides the two point correlation function at super-horizon scales (see e.g.~\cite{Riotto:2002yw})
\begin{equation}
\left\langle \pi_{\bf k} \pi_{\bf k'} \right\rangle = 
\frac{a^2 H_I^2}{2 k^3} \delta^{(3)} \left( {\bf k} + {\bf k'} \right) \equiv \frac{2 \pi^2}{k^3} {\cal P}_\pi \left( k \right) \;,
\label{defscps}
\end{equation}
where the final expression is the standard definition of the power spectrum (in the following, we use an identical relation for the power spectra of other field variables). From the rescalings~(\ref{relpphi}) and (\ref{rel_aL}) we obtain the super-horizon power spectra~\cite{Graham:2015rva}: 
\begin{equation}
{\cal P}^{(0)}_{\varphi} \left( k \right)
= \frac{H_I^2}{4 \pi^2\,M^2} \;\;\;\Rightarrow\;\;\; 
{\cal P}^{(0)}_{A_L} \left( k \right) =\frac{k^2}{M^2}\,
\frac{H_I^2}{4 \pi^2}\,, 
\label{amspLM}
\end{equation}
where the suffix (0) has been added to indicate that these quantities are constant, and they provide initial conditions for the mode evolution during the radiation dominated era. 
The primordial spectrum for the longitudinal scalar $\varphi$ is scale invariant, where the amplitude is
controlled by the vector mass and the inflationary Hubble parameter. In contrast the primordial spectrum of the longitudinal vector $A_L$ grows as $k^2$ during inflation \cite{Graham:2015rva} -- recall the relation \eqref{rel_aL}. This scale dependence allows us to avoid CMB isocurvature constraints at (large) CMB scales. However, during radiation domination, the spectrum is not a monotonic function of $k$ and instead develops a peak, as we will review in Section
\ref{sec_evRD}. 

\medskip

While Eq. \eqref{amspLM} is the result  of the minimal scenario
proposed in \cite{Graham:2015rva}, it is possible  to generalise the original idea along various directions. Beyond their intrinsic interest, generalisations are particularly relevant for the study of induced gravitational waves -- see Sections \ref{sec_SGWB} and \ref{sec_pheno}. In fact, only extensions of \cite{Graham:2015rva} can produce a gravitational wave spectrum with a sufficiently large amplitude to be detectable by gravitational wave experiments.

In the literature, attempts to generalise  \cite{Graham:2015rva} move along different directions, by

\begin{enumerate}
\item Including reheating
effects, or changes in
cosmological
history between  the phases of
de Sitter inflation and the
onset of radiation domination, see e.g. \cite{Ema:2019yrd,Ahmed:2020fhc,Kolb:2020fwh}.
 In this case, the evolution of mode functions depend on the early cosmological evolution, which influences 
 the  amplitude and scale dependence of the vector energy density during radiation domination. 
\item Considering
non minimal couplings
to gravity during inflation, see e.g. 
\cite{Ozsoy:2023gnl,Capanelli:2024rlk}, as well as the review \cite{Kolb:2023ydq}. For example, a ghost-free vector coupling with the Einstein tensor  as $\alpha\, A^\mu A^\nu\,G_{\mu\nu}$, controlled by a constant   $\alpha$, leads to a longitudinal scalar spectrum amplitude
${\cal P}^{(0)}_{\varphi}
= {H_I^2}/{\left(4 \pi^2\,\left(M^2 +\alpha^2 \,H_I^2\right)\right)} $ during inflation \cite{Ozsoy:2023gnl}.
\item Envisaging mechanisms as phase transitions that change the value of the vector mass during  inflation, or immediately after such phase; see e.g. \cite{Salehian:2020asa} for an explicit construction. If the change of mass happens while the mass is much smaller than the Hubble rate and the physical momentum of a given mode, the vector mode function (and its time derivative) do not change in any appreciable way. However, as we show below, the vector energy density and its coupling to the gravitational waves increase if the vector mass increases, and, as seen for instance from eqs. (\ref{veda}) and (\ref{intOGW}), a change of the mass has the exact same effect as a change in the amplitude of the primordial
longitudinal spectrum whilst the mode is well outside the horizon. So, we can mathematically treat this case analogously to the two previous ones.
\end{enumerate}

We stress that extensions of the original model \cite{Graham:2015rva} are expected to introduce some degree of specific scale dependence in the spectrum of longitudinal vector modes. However, since our focus is not on specific model building, we assume - both for simplicity and phenomenological reasons - that such generalisations only modify overall normalisation of the longitudinal mode spectrum at the onset of radiation domination.  
Namely, we assume that Eq. \eqref{amspLM} is generalised to the formula
\bea
\label{pphiin2}
{\cal P}^{(0)}_{\varphi}(k)
\,=\,\sigma_0\,\frac{H_I^2}{4 \pi^2\,M^2}
\,,
\eea
where $\sigma_0$ is a constant dependent on specific scenarios that go beyond \cite{Graham:2015rva}. For our purposes, we assume $\sigma_0$ is a free parameter. As explained above, it will play an important role for the considerations in Section \ref{sec_pheno}.

\subsection{Evolution during
radiation domination}
\label{sec_evRD}

We take Eq. \eqref{pphiin2} as reference for the initial primordial spectrum and evolve it during radiation domination (RD). 
This is done by using transfer functions and expressing Eq. \eqref{int_trf} during RD ($\tau\ge0$)
as
\be
\label{int_trf2}
\varphi_{\bf  k} (\tau)\,=\,T(k \tau)\,\varphi_{\bf  k}^{(0)}\,,
\ee
with $\varphi_{\bf  k}^{(0)}$ the initial value of the longitudinal scalar leading to
the   spectrum in Eq. \eqref{pphiin2}.
We follow the approach~\footnote{GT thanks Ogan \"Ozsoy for collaborating 
on \cite{Ozsoy:2023gnl} and for providing key insights in the treatment of transfer functions.
} outlined in \cite{Ozsoy:2023gnl} to handle the transfer functions.

\smallskip
\begin{figure}[t!]
    \centering
    \includegraphics[width=0.44\linewidth]{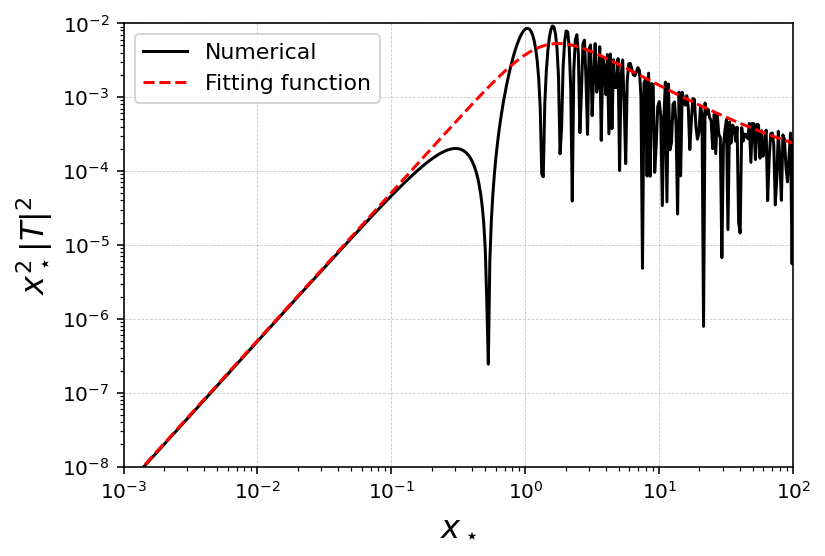}\includegraphics[width=0.44\linewidth]{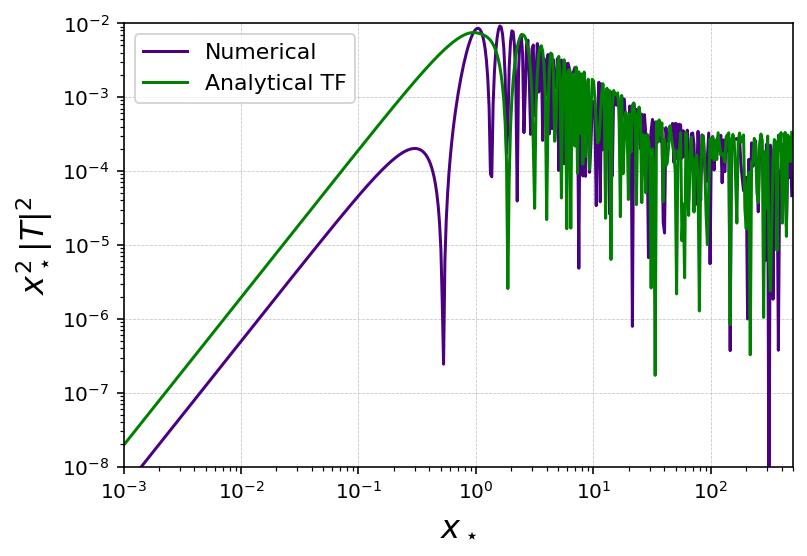}
    \caption{\small Representation of the longitudinal mode spectrum $({2 \pi M})^2/({\sqrt{\sigma_0}\,k_\star\,H_I})^2\, {\cal P}_{A_L}\,=\,x_\star^2\,|T|^2$.  {\bf Left Panel:} Numerical approach.      We evaluate this quantity at time $y_{fin}=55$, while $y_{in}=10^{-3}$. In red dashed the fitting function \eqref{psfit}. See Section \ref{sec_evRD}. {\bf Right Panel:} Numerical versus  analytical
    spectrum using the analytical transfer
    functions in RD as developed in Section \ref{subsec_ant}.} 
    \label{fig:transfer}
\end{figure}

Using Eq.
 \eqref{eveqf}, we find that 
the transfer
function of Eq. \eqref{int_trf}  obeys the following evolution equation during RD:
\bea
\label{evTfv}
T''(x)+\frac{2}{x}\,\left(1+\frac{a^2\,M^2}{k^2}\right)^{-1}
T'(x)+\left(1+\frac{a^2\,M^2}{k^2}\right)\,T(x)
\,=\,0\,,
\eea
where prime denotes differentiation with respect to $x \equiv k \tau$, and we impose
$T\to1$ at superhorizon scales $x\ll1$. The mass $M$
indicates the vector mass during RD. 
In this era, $a H =1/\tau$. The Hubble rate decreases with respect to the inflationary value $H_I$, $a_*$ denotes the value of the scale factor when the rate becomes equal to the vector mass,  
\be
\label{def_hst}
H_\star\, \equiv \,
H (a_\star)\,=\,M
\ee
It is convenient to define the pivot scale
\be
\label{def_kst}
k_\star \equiv a_\star M\,.
\ee
As we will review next, $k_\star$ marks a turning point in the slope of the longitudinal vector spectrum profile during RD, where the slope transitions from increasing to decreasing as a function of momentum. We introduce the notation
\bea
x&\equiv&k \tau
\hskip0.5cm;\hskip0.5cm
y\,\equiv\,\frac{x}{x_\star}
\hskip0.5cm;\hskip0.5cm
x_\star\,\equiv\,\frac{k}{k_\star}
\,.
\label{def_ys}
\eea
Recall that $a\propto \tau$ during RD, hence we can write $y=a/a_\star$. Since $d y/d x\,=\,1/x_\star$, in terms of these variables the evolution of the transfer function reads 
\be
\label{eveq_tr2}
\frac{d^2 T(y)}{d y^2} +\frac{2\,(1+y^2/x_\star^2)^{-1}}{y}\,
\,\frac{d T(y)}{d y} +(x_\star^2+y^2)\,T(y)\,=\,0\,.
\ee
This is numerically integrated for a set of different values $x_\star= k/k_\star$, which correspond to different comoving momenta normalised to the pivot momentum $k_*$. For each integration, we impose the initial conditions $T(y_{\rm in})=1$, $\frac{d T(y_{\rm in})}{d y} =0$, with~\footnote{Since $H \propto t^{-1} m\propto a^{-2}$ during RD, the choice $y_{\rm in} \equiv \sqrt{M/H_I}$ corresponds to $H_{\rm in} = H_* \left( a_* / a_{\rm in} \right)^2 = H_I$. Namely, the variable $y$ evaluates to $y_{\rm in}$ at the onset of RD, which in this computation is assumed to occur right after the end of inflation.} $y_{\rm in} \equiv a_{\rm in}/a_{\star}\, \equiv \,\sqrt{M/H_I}$. The result is evaluated at a sufficiently late time, when $y_f = \frac{a_f}{a_*} \, {\rm e}^4$, when the longitudinal spectrum stabilises its shape as a function of frequency.

The resulting spectrum of longitudinal modes $A_L$ during RD at time $y_f$ is
\be
{\cal P}_{A_L}(k, y_f)\,=\,
\left( \frac{\sqrt{\sigma_0}\,k_\star\,H_I}{2 \pi M}\right)^2
\,\frac{k^2}{k_\star^2}
\,\left|
T(x_\star, y_f) \right|^2
\,.
\ee
The left panel in Fig \ref{fig:transfer} is a numerical evaluation of this spectrum, where it is normalised versus $({\sqrt{\sigma_0}\,k_\star\,H_I})^2/({2 \pi M})^2$, with the $\sigma_0$ the factor defined in Eq. \eqref{pphiin2}. Notice that, as anticipated, the spectrum changes slope at around $x_\star =1$, i.e. for scales $k\sim k_\star$. 
In the same figure we also represent an analytical fitting function
\bea
\frac{({2 \pi M})^2}{({\sqrt{\sigma_0}\,k_\star\,H_I})^2}\, 
{\cal P}_{A_L}(k, y_{\rm fin})&=&
x_\star^2 |T(x_\star, y_{\rm fin})|^2
\nonumber
\\
&
=&
 10^{-2}
\left( \frac{x_\star}{1.4}\right)^2
\left[1+ \left( \frac{x_\star}{1.4} \right)^3 \right]^{-1}+10^{-4}
\left( \frac{x_\star}{10}\right)^2 
\left[1+ \left( \frac{x_\star}{10} \right)^2 \right]^{-1}\,,
\nonumber\\
\label{psfit}
\eea
indicating that the spectrum
increases as $(k/k_\star)^2$ from large to
small scales, reaching a maximum at $k\sim k_\star$, then decreases as $(k/k_\star)^{-1}$, and finally stabilises to a constant value at very small scales. This behaviour is consistent with the results of \cite{Graham:2015rva}.

\smallskip
Numerical analysis demonstrates that the longitudinal vector spectrum has a peak at small scales during RD --
a noteworthy feature that can produce induced GW at second order in fluctuations. However, beyond numerics, it is desirable to have better analytical control over the transfer function for the longitudinal scalar during RD. Which we address in the next session.

\subsection{New analytic approach to the transfer function}
\label{subsec_ant}

\smallskip
\begin{figure}[t!]
    \centering
    \includegraphics[width=0.46\linewidth]{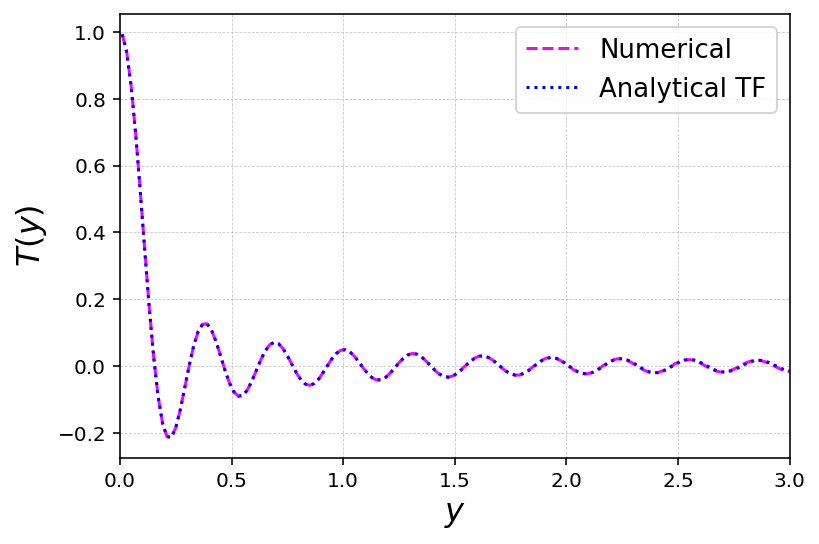}
     \includegraphics[width=0.46\linewidth]{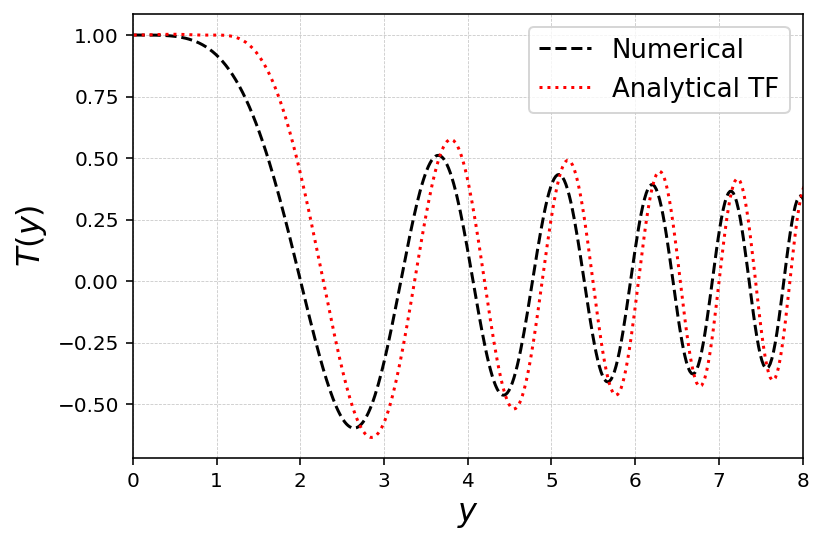}
    \caption{\small
{\bf Left panel} Numerical solution
of transfer function in the small $y/x_\star$  regime (magenta), versus the analytical solution \eqref{solT1} (blue). We choose $x_\star=20.3$. {\bf Right panel} Numerical solution
of transfer function in the large $y/x_\star$  regime (black), versus the analytical solution \eqref{solT2} (red). We  choose $x_\star=0.01$.}
    \label{fig:solstransfer}
\end{figure}
The evolution equation \eqref{eveq_tr2} for the transfer functions
during RD has analytical  solutions in the regimes~\footnote{Recall 
the definitions of $y$ and $x_\star$ in Eq. \eqref{def_ys}.} of small and large $y$.

\smallskip
\noindent
{\bf Small $y/x_\star$:} The solution of \eqref{eveq_tr2}, using the same definition of $y_{\rm in}$ and the appropriate initial conditions, is given by
\be
\label{solT1}
T(y)\,=\,
\frac{\sin \left[ x_\star \left( y -y_{\rm in} \right) \right]}{x_\star y} + \frac{y_{\rm in} \cos \left[ x_\star \left( y-y_{\rm in} \right) \right]}{y}\,,
\ee

\smallskip
\noindent
{\bf Large $y/x_\star$:}  The solution is
\be
\label{solT2}
T(y)\,=\,c_1 \, D_{-1/2} \left[ (1+i) y \right]
+
c_2 \, D_{-1/2}\left[ -(1+i) y \right]\,,
\ee
with $D_\nu [z]$ the
parabolic cylinder function, and the constants
$c_{1,2}$ are fixed
by boundary conditions.

\smallskip
Using  asymptotic formulas for the cylinder functions, we determine a convenient approximation, which is reasonably accurate for all values of $y$ and $x_\star$. It is defined in two  regimes of $x_\star$ (from now
on we set for simplicity $y_{\rm in}=0$):

\begin{itemize}
\item[1.] Range $x_* \ll 1$
\begin{eqnarray}
T_{\rm early} &=& \frac{\sin \left( x_* y \right)}{x_* y} \;\;\; \hskip4.5cm {\rm valid \; for \; } y \ll 1  
\,,
\nonumber\\ 
T^{(A)}_{\rm late} &=&  \frac{1}{\sqrt{y}} \left[ c_1^A \, \cos \left( \frac{y^2}{2} \right) + c_2^A \, \sin \left( \frac{y^2}{2} \right) \right] \;\;\; \hskip0.2cm {\rm valid \; for \; } y \gg 1\,,
\label{eq_trxs}
\end{eqnarray}
with

\begin{eqnarray}
c_1^A &=& - \sin\left( {1}/{2}\right) \; \cos x_* + \left(  \cos\left( {1}/{2}\right) + \frac{1}{2} \sin \left( {1}/{2}\right)\right) \, \frac{\sin x_*}{x_*} \,, \nonumber\\ 
c_2^A &=& \cos\left( {1}/{2}\right) \; \cos x_* + \left(  \sin\left( {1}/{2}\right) - \frac{1}{2} \cos\left( {1}/{2}\right) \right) \, \frac{\sin x_*}{x_*}\,.
\end{eqnarray}

\item[2.] Range $x_* \gg 1$

\begin{eqnarray}
T_{\rm early} &=& \frac{\sin \left( x_* y \right)}{x_* y} \;\;\;  \hskip4.5cm {\rm valid \; for \; } y \ll x_\star 
\,,
\nonumber\\ 
T^{(B)}_{\rm late} &=&  \frac{1}{\sqrt{y}} \left[ c_1^B \, \cos \left( \frac{y^2}{2} \right) + c_2^B \, \sin \left( \frac{y^2}{2} \right) \right] \;\;\;  \hskip0.2cm  {\rm valid \; for \; } y \gg x_\star 
\,,
\label{eq_trxb}
\end{eqnarray}
with 
\begin{eqnarray}
c_1^B &\simeq& \frac{\sin \left( {x_*^2}/{2} \right)}{x_*^{3/2}} \left( 1 + \frac{\sin x_*^2}{2 x_*^2} \right) 
\,,
\nonumber\\ 
c_2^B &\simeq& \frac{\cos \left( {x_*^2}/{2} \right)}{x_*^{3/2}} \left( 1 - \frac{\sin x_*^2}{2 x_*^2} \right)
\,.
\end{eqnarray}

\end{itemize}

We extend the region of validity of the above expressions (\ref{eq_trxs}) and (\ref{eq_trxb}) to include also the transition regions at $y=1$ and $x_* = 1 ,\, y$, by replacing the strong inequalities $\ll$ and $\gg$ with, respectively, $\leq$ and $\geq$. The above choice of the $y-$independent coefficients $c_{1,2}^{A/B}$ makes this analytic approximation for $T$, as well as its derivatve $\frac{\partial T}{\partial y}$ continuous across the transition regions. In Fig \ref{fig:solstransfer} we compare our analytical approximation
of the transfer functions against
numerics for two representative values of $x_\star$. Numerical checks show 
 that our previous
formulas \eqref{eq_trxs} and \eqref{eq_trxb} give reasonably accurate results also  for intermediate values of $x_\star$.  In Fig \ref{fig:transfer}, right panel, we compare the longitudinal mode spectrum as function of scale $x_\star=k/k_\star$,
computed with our analytical transfer functions against numerics. The profile (and the maximum) of the spectrum are captured quite well -- in particular the position and amplitude of the peak,  the spectral slope,
as well as the oscillatory behaviour towards small scales.

\smallskip
In fact, in  deriving Eqs  \eqref{eq_trxs} and \eqref{eq_trxb} we make an effort to catch the oscillatory behaviour of the transfer functions. The latter  plays an important role in the
 convolution integrals of Section \ref{sec_SGWB} for the computation of the induced GW spectrum. As the present work was being completed, similar examples of transfer functions have appeared in \cite{Gorji:2025tos} in a related (but not identical) context.

\subsection{The  energy density in the longitudinal vector mode}

The computation of the energy density $\rho_{A_L}$ stored in the longitudinal vector field is 
a useful first application of the analytic formulas derived in Section \ref{subsec_ant}. 
The quantity $\rho_{A_L}$ is extracted from the time-time component of the vector energy momentum tensor. It is given by \cite{Graham:2015rva}:
\bea
\rho_{A_L}&=&\frac{M^2}{2 a^2}\,\int d \ln k\,\left[ \frac{ {\cal P}_{\left(\partial_\tau A_L \right)}}{k^2+a^2\,M^2} +
{\cal P}_{A_L}
\right]\,,
\label{veda}
\eea
with $M$ the vector mass during RD. 
Substituting  expression \eqref{amspLM} 
for the primordial longitudinal vector spectrum (including
the $\sigma_0$ factor of Eq. \eqref{pphiin2}), and using the transfer functions,  this expression becomes
\bea
\rho_{A_L}
&=&
\frac{\sigma_0\,H_I^2}{8 \pi^2\,a^2}
\frac{a_\star \,k_\star^2}{a} \left[
\int d \ln k\,
\left(\frac{k^2\,a}{a_\star \,k_\star^2(k^2+a^2 M^2)} |\partial_\tau T|^2+\frac{a\,k^2}{a_\star \,k_\star^2} |T|^2 \right)\right]\,.
\label{vedb}
\eea
In passing from Eq. \eqref{veda} to Eq. \eqref{vedb}, we acquire a factor proportional to $4 \pi^2\,M^2 {\cal P}_\varphi^{(0)}/H_I^2$.
In the original scenario \cite{Graham:2015rva},
this factor is equal to one (see Eq. \eqref{amspLM}). Due to the fact that the primordial inflationary spectrum is proportional to the inverse of the square of the vector mass during inflation (recall the arguments leading to Eq. \eqref{amspLM}); when considering more general models, for example where the vector mass during inflation has value $m_{\rm inf}$ (which is different from the RD vector mass $M$). This factor leads to an effective overall constant $\sigma_0 \,=\,M^2/m_{\rm inf}^2$ in Eq. 
\eqref{vedb}. Effectively, in this particular case, this phenomenon is captured by the factor $\sigma_0$
in Eq. \eqref{pphiin2}.
More generally,
in what follows -- without focussing
on specific scenarios -- we keep the parameter
$\sigma_0$ as free constant in Eq. \eqref{vedb}.

We focus on the dimensionless integral inside
the square bracket in  formula \eqref{vedb}
\bea
{\cal I}_{\rho}(y)&=&\int d \ln k\,
\left[\frac{k^2\,a}{a_\star \,k_\star^2(k^2+a^2 M^2)} |\partial_\tau T|^2+\frac{a\,k^2}{a_\star \,k_\star^2} |T|^2 \right]
\\
&=&y\,\int d \ln x_\star\,
\left[\frac{x_\star^2}{x_\star^2+y^2} |\partial_y T|^2+x_\star^2 |T|^2 \right]\,,
\eea
where in the second line we pass to the dimensionless `time'  coordinate $y$ introduced in Eq. \eqref{def_ys}.
We evaluate this integral at late times, which we denote with $y=y_{\rm end}$, making use of the analytical transfer functions determined in Section \ref{subsec_ant}. The quantity $y_{\rm end}$ is chosen to be sufficiently large so that the following integration converges to a stable value. Specifically, given the structure of our analytic solutions for the transfer functions, see Section \ref{subsec_ant}, we conveniently  split the integral in two pieces, for $y<1$ and $y>1$: 
\bea
{\cal I}_\rho(y_{\rm end})&=&y_{\rm end}\,\int_0^1 \frac{d  x_\star}{x_\star}\,
\,
\left[\frac{x_\star^2}{x_\star^2+y_{\rm end}^2} |\partial_y T^{(A)}_{\rm late}|^2+x_\star^2 |T^{(A)}_{\rm late}|^2 \right]
\nonumber
\\
&&+y_{\rm end}\,\int_{1}^{y_{\rm end}} \frac{d  x_\star}{x_\star}\,
\left[\frac{x_\star^2}{x_\star^2+y_{\rm end}^2} |\partial_y T^{(B)}_{\rm late}|^2+x_\star^2 |T^{(B)}_{\rm late}|^2 \right]
\,.
\label{int2pie}
\eea

Consider the  integral in the first line of the Eq. \eqref{int2pie}. We
first take the limit of large $y_{\rm end}$ inside the integrand function. The result is easily numerically integrated, and gives the value 0.475883.
For the  integral in the second line of  Eq. \eqref{int2pie} -- since the 
quantity $y_{\rm end}$ appears both in the integral extreme and
in the integrand -- we find convenient
to first rescale the integration variable from $x_*$ to $z \equiv \frac{x_*}{y_{\rm end}}$. In this new variable, the lower extreme of integration $\frac{1}{y_{\rm end}}$ is small and can be approximated with zero. Morevoer, in the integrand function that results from this change of variable, there are terms that exhibit rapid oscillations in the $0 \leq z \leq 1$ domain, which are weighted by inverse powers
of $y_{\rm end}$. We can neglect those contributions, since they average to nearly zero. Beside these contributions, the integrand contains a ratio
between two polynomials in the variable $x_\star$, which once integrated gives 1.025 in the limit of large $y_{\rm end}$. The
sum of the two pieces we computed is 
\be
{\cal I}_\rho\,=\,
1.50088\,\simeq\,\frac32\,,
\label{resone}
\ee
a number 
 of order one.  With this result, and recalling the definition of $H_\star$ in Eq. \eqref{def_hst},  we obtain the following expression  
 for the longitudinal vector energy density evaluated today
 \be
 \rho_{A_L}\,=\,\frac{3 {\sigma}_0}{2}\,\frac{H_I^2\,M^2}{8 \pi^2}\,\left(
 \frac{H_{\rm eq}}{M}
 \right)^{3/2}\,,
 \ee
 where we take into account how the energy density 
 scales with the universe evolution.
 The present-day
  DM energy density is well approximated by the formula 
 $\rho_{\rm DM}\,=\,3 \,H_{\rm eq}^2\,M_{\rm Pl}^2/2$. Using
 the value 
 $H_{\rm eq}= 3\times 10^{-28}$
 eV, we  compute the ratio of these two quantities 
 \be
 \label{eq_ratioen}
 \frac{\rho_{A_L}}{\rho_{\rm DM}}\,=\,{\sigma}_0\,
\left( \frac{M}{0.6\times 10^{-6} \,{\rm eV}}\right)^{1/2}\,\left(\frac{H_I}{10^{14}\,{\rm GeV}}\right)^2 \,,
 \ee
 giving a result that is in reasonable agreement~\footnote{There are order-one differences between our Eq. \eqref{eq_ratioen} and  the analogue  formula in \cite{Graham:2015rva}. We interpret the difference as due to our approximations in deriving
 the analytical transfer
 functions.} with \cite{Graham:2015rva}. We will make use of expression \eqref{eq_ratioen} in Section \ref{sec_pheno}.

\section{The induced gravitational wave spectrum}
\label{sec_SGWB}

The vector dark matter model reviewed in the previous
section leads to a longitudinal vector spectrum
 enhanced at scales smaller than the CMB ones, with a peak at  $k \simeq k_{\star}\,\equiv \,a_\star\,M$, see Fig \ref{fig:transfer}. 
The fluctuations of the longitudinal vector produce tensor metric perturbations (gravitational waves) at second order, in an amount quadratically proportional to their power spectrum. This leads us to investigate whether the amplifcation of the longitudinal vector in this model can result in a visible nduced GW signal. Such phenomenon is analog to what happens in models  producing primordial
black holes through an enhancement of curvature fluctuations, which
induce a SGWB at second order in perturbations (see e.g. \cite{Domenech:2021ztg} for a comprehensive review). 

Interestingly, in our context we do not need to consider deviations from standard slow roll expansion during inflation, nor the production of primordial black holes. The induced GW background is an inevitable consequence of the shape of the longitudinal vector spectrum associated
with the production of vector dark matter. 
We investigate this subject following the approach  developed in \cite{Ananda:2006af,Baumann:2007zm}, generalising it to the case of non-adiabatic fluctuations. In the next section we discuss prospects of detection. We find that the system we consider produces an SGWB at and below nano-Hertz frequencies.

\smallskip

We consider mode evolution during radiation domination. Inflationary 
GW are  transverse-traceless, spin-2 fluctuations around a  FLRW background metric.  They 
satisfy the evolution equation (using the notation of \cite{Baumann:2007zm})
\be\label{eq_strheq}
h_{ij}''+2 {\cal H}\,h_{ij}'-\nabla^2 h_{ij}
\,=\,{\cal T}_{ij}^{\,\,l m}S_{l m}\,,
\ee
where $S_{ij}$ is the longitudinal vector source, and
${\cal T}_{ij}^{\,\,l m}$ is the projection tensor selecting its transverse-traceless part. The GW source is extracted from the spatial components of the vector energy-momentum tensor. In our context
 it reads
  \bea
S_{ij}(\tau, \bf x)&=&2 M^2\,\partial_i \varphi \,
\partial_j \varphi-\frac{2}{a^2(\tau)} \partial_i A_0\,\partial_j A_0-\frac{2}{a^2(\tau)} \partial_i \varphi'\,\partial_j  \varphi'
\nonumber
\\
&&+\frac{2}{a^2(\tau)} \left( \partial_i A_0\, \partial_j \varphi'+
\partial_j A_0\, \partial_i \varphi'
 \right)\,.
\eea
It depends on the vector mass $M$ during RD,
the scalar longitudinal mode $\varphi$, and the time-component $A^0$ of the vector. Hence, at second order in fluctuations, the vector sources GW, which are expanded in Fourier modes as
\bea
h_{ij} (\tau, {\bf x})&=&
\sum_{\lambda}
\,\int \frac{d^3 {\bf k}}{(2 \pi)^{3/2}} \,e^{i {\bf k}  \cdot {\bf x} }\,
{\bf e}^{(\lambda)}_{ij}({\bf k})\,h_{\bf k}^{(\lambda)} \,,
\eea
where the sum over $\lambda$ is a sum over the two GW polarizations $(+,\,\times)$. The
polarisation tensors ${\bf e}^{(\lambda)}_{ij}$ are normalised as $\sum_{ij} {\bf e}^{(\lambda)}_{ij}
{\bf e}^{(\lambda')}_{ij}\,=\,\delta^{\lambda \lambda'}$, and they can also be used to perform the transverse-traceless projection tensor of Eq.\eqref{eq_strheq}. The Fourier transform of the source
on the right-hand side of Eq. \eqref{eq_strheq} 
reads~\cite{Baumann:2007zm}
\bea
{\cal T}_{ij}^{\,\,\,lm} S_{l m}(\tau, {\bf x})
&=& \sum_\lambda\,\int
 \frac{d^3 {\bf k}}{(2 \pi)^{3/2}} \,e^{i {\bf k} \cdot {\bf x} }\,
\left[ 
{\bf e}^{(\lambda)}_{ij}({\bf k})
{\bf e}^{(\lambda)\,lm}({\bf k})
\right]\,S_{l m} ({\bf k})\,,
\eea
with
\be
S_{l m} ({\bf k})\,=\,\int \frac{d^3 {\bf x'}}{(2 \pi)^{3/2}} \,e^{-i {\bf k} \cdot {\bf x}' }\,S_{lm} (\bf x')
\,.
\ee
Hence, the GW evolution equation \eqref{eq_strheq} reads in Fourier space
\be
\label{eq_gwfsv}
h^{(\lambda)\,''}_{\bf k}+2 {\cal H}\,h^{(\lambda)\,'}_{\bf k}+k^2 h_{\bf k}^{(\lambda)}
\,=\,S_{\bf k}^{(\lambda)}(\tau)\,, 
\ee
with the source Fourier component given by the convolution integral
\bea
\hskip-0.1cm
S^{(\lambda)}_{\bf k}&=&
2 M^2 \int
\frac{{\texttt e}^{(\lambda)}({\bf k}, \tilde{\bf k})\,\,d^3 {\bf \tilde k}}{(2 \pi)^{3/2}}\, 
\,
\nonumber
\\
&&
\times 
\Big[
\varphi_{\bf \tilde k}   \varphi_{\bf k -\bf \tilde k} 
-
\frac{1}{a^2\,M^2} 
\left(
 A_{0 ,{\bf \tilde k}   }\,A_{0,{\bf k -\bf \tilde k}}  
 -
  \varphi'_{\bf \tilde k}   \varphi'_{\bf k -\bf \tilde k} 
  -A_{0 , \, {\bf \tilde k}   }
\,  \varphi'_{\bf k -\bf \tilde k} +
A_{0 ,\, {\bf k -\bf \tilde k}   }
\,  \varphi'_{\bf k }
  \right)
  \Big]\,.
\label{sourfour}
\eea
where we introduced the combination 
\be
{\texttt e}^{(\lambda)}({\bf k}, \tilde{\bf k})
\,\equiv\,{\bf e}^{(\lambda)}_{ij}( {\bf k})\,\tilde k^i \tilde k^j 
\label{intrsce}
\,.
\ee
It is convenient to make use of the constraint equation \eqref{rel_aze} relating the vector time component $A_0$ and the longitudinal scalar $\varphi$: $A_{0 \, {\bf  k}   }\,=\,{\left(k^2 \, \varphi'_{\bf  k} \right)}{/(k^2+M^2 a^2)}$. Substituting such condition into Eq. \eqref{sourfour}, we find 
\bea
S^{(\lambda)}
_{\bf k}&=&2\,M^2\,
\int
\frac{{\texttt e}^{(\lambda)}({\bf k}, \tilde{\bf k})\,d^3 {\bf \tilde k}}{(2 \pi)^{3/2}}
\,\left[\varphi_{\bf \tilde k}   \varphi_{\bf k -\bf \tilde k} -
\frac{a^2\,M^2\,\varphi'_{\bf \tilde k}   \varphi'_{\bf k -\bf \tilde k} }{\left( \tilde k^2+a^2 M^2 \right)\left( | {\bf k}-{\bf \tilde k}|^2+a^2 M^2 \right)} 
\right]\,.
\label{sourfour2}
\eea
Notice that the argument of the previous integral contains non-local contributions proportional 
to the inverse of momenta. They originate from the procedure of integrating out the time-like
component of the vector, leading to non-canonical kinetic terms for the longitudinal mode. 
To proceed, we make use of the results of Section \ref{sec_rev}: the  Fourier component of the longitudinal scalar $\varphi$ during radiation-domination (RD) can be expressed in terms of the transfer function
\be
\label{int_trf}
\varphi_{\bf  k} (\tau)\,=\,T(k \tau)\,\varphi_{\bf  k}^{(0)}\,,
\ee
where the transfer function $T(k \tau)$ during  RD is discussed in Section \ref{sec_evRD}, and $\varphi_{\bf  k}^{(0)}$ corresponds
to the scalar primordial  Fourier mode  evaluated
at the end of inflation. The source Fourier mode \eqref{sourfour2}  corresponds to a convolution
over primordial longitudinal modes 
\bea
S^{(\lambda)}_{\bf k}(\tau)& =&2 M^2\,
\int
\frac{{\texttt e}^{(\lambda)}({\bf k}, \tilde{\bf k})\,d^3 {\bf \tilde k}}{(2 \pi)^{3/2}}\,\beta(\tau, k, \tilde k)\,
\varphi_{\bf \tilde k}^{(0)} \varphi_{\bf  k-\bf \tilde k}^{(0)}\,,
\eea
weighted by a combination  of transfer functions which we denote by
\bea
\label{defff}
\beta(\tau, {\bf k}, {\bf \tilde k})&\equiv& 
T(\tilde k \tau)  T(| {\bf k}-{\bf \tilde k}| \tau)-\frac{a^2\,M^2\,\tilde k\,| {\bf k}-{\bf \tilde k}|\, T'(\tilde k \tau)  T'(| {\bf k}-{\bf \tilde k}| \tau) }{\left( \tilde k^2+a^2 M^2 \right)\left( | {\bf k}-{\bf \tilde k}|^2+a^2 M^2 \right)} 
\,, 
\eea
where a prime indicate a derivative along the argument. We note that the combination $\beta$ 
is invariant under the exchange $\tilde k \leftrightarrow | {\bf k}-{\bf \tilde k}|$. 

\smallskip
After characterising its source, we can express
the formal solution of the GW evolution Eq. \eqref{eq_gwfsv} during RD:
\be
h^{(\lambda)}_{\bf k}(\tau)\,=\,\frac{1}{a(\tau)}\,\int d \tau' g_{k}(\tau, \tau') \left[ a(\tau')\,S^{(\lambda)}_{\bf k}(\tau')
\right]\,,
\ee
where the $g_k$ is the Green function (or, more appropriately, the non-distributional contribution to the Green function). During radiation domination: 
\be
g_{k}(\tau, \tau')\,=\,\frac{1}{k} \left[ \sin{(k \tau)} \cos{(k \tau')}-\sin{(k \tau')}
\cos{(k \tau)}
\right]\,.
\ee
We consider the tensor spectrum as the sum over the two polarizations
\be
\sum_\lambda \langle 
{h^{(\lambda)}_{\bf k}(\tau)}
h^{(\lambda)}_{\bf q}(\tau)
\rangle
\,=\,
(2 \pi)^3\,\delta^{(3)}({\bf k}+{\bf q})\,\frac{4 \pi^2}{k^3}\,{\cal P}_{h}(k)
\,.
\ee
which we rewrite as (the prime in the ensemble average indicates the correlation without the $\delta$-function associated with momentum conservation)
\bea
{\cal P}_h(\tau, k)
&=&\frac12\,\frac{k^3}{2 \pi^2}
\,\sum_\lambda
\langle
h^{(\lambda)}_{\bf k}(\tau) 
h^{(\lambda)}_{\bf q}(\tau) 
\rangle'
\\
&=&\frac{ k^3}{4 \pi^2} 
\frac{\sum_\lambda}{a^2(\tau)}
\int d \tau_1 d \tau_2\,g_{ k}(\tau, \tau_1)\,g_{ q}(\tau, \tau_2)\,a(\tau_1) a(\tau_2)
\langle
S^{(\lambda)}_{\bf k}(\tau_1)
S^{(\lambda)}_{\bf q}(\tau_2)
\rangle \,,  
\label{eq_ts2s}
\eea
where the two-point function of the source $S_{\bf k}$ in Fourier space reads 
\bea
\langle S^{(\lambda)}_{\bf k}(\tau_1)
S^{(\lambda')}_{\bf q}(\tau_2)
\rangle
&=&4 M^4\,
\int
\frac{d^3 {\bf \tilde k}}{(2 \pi)^{3/2}}\,\frac{d^3 {\bf \tilde q}}{(2 \pi)^{3/2}}\,{\texttt e}^{(\lambda)}({\bf k}, \tilde{\bf k})\,{\texttt e}^{(\lambda')}({\bf q}, \tilde{\bf q})
\nonumber
\\
&&
\hskip0.7cm
\times
\,\beta(\tau_1, {\bf k}, {\bf \tilde k})\,\beta(\tau_2, {\bf q}, {\bf \tilde q})\,
\langle
\varphi_{\bf \tilde k}^{(0)} \varphi_{\bf  k-\bf \tilde k}^{(0)}
\varphi_{\bf \tilde q}^{(0)} \varphi_{\bf  q-\bf \tilde q}^{(0)}
\rangle\,.
\label{eq_stpf}
\eea
Expanding the scalar four-point function in
Eq. \eqref{eq_stpf} by means of Wick's theorem, and  using the definition \eqref{defscps} of primordial
longitudinal   scalar spectrum, we  find
\bea
\sum_\lambda
\langle S^{(\lambda)}_{\bf k}(\tau_1)
S^{(\lambda)}_{\bf q}(\tau_2)
\rangle
&=&8 M^4\,(2 \pi^2)^2\,\delta^{(3)}({\bf k}+{\bf q})
\int
\frac{d^3 {\bf \tilde k}}{(2 \pi)^{3}}\,\left(\sum_\lambda
{\texttt e}^{(\lambda)}({\bf k}, \tilde{\bf k})  {\texttt e}^{(\lambda)}({\bf k}, \tilde{\bf k}) \right)
\,
\nonumber
\\
&& \hskip3cm \times \,\beta(\tau_1, {\bf k}, {\bf \tilde k})\,\beta(\tau_2, {\bf k}, {\bf \tilde k})\, \frac{{\cal P}^{(0)}_\varphi (\tilde k) }{\tilde k^3}
 \frac{{\cal P}^{(0)}_\varphi (|{\bf k}- {\bf \tilde k}|) }{|{\bf k}- {\bf \tilde k}|^3}
 \nonumber
\\
&=&8 \pi^2 M^4\,
\delta^{(3)}({\bf k}+{\bf q})
\int_0^{\infty}
d  \tilde k  \, \int_{-1}^1 d \mu\,\frac{\tilde k^3 (1-\mu^2)^2}{|{\bf k}- {\bf \tilde k}|^3}
\,\beta(\tau_1, {\bf k}, {\bf \tilde k})\,\beta(\tau_2, {\bf k}, {\bf \tilde k})\,
\nonumber
\\
&& \hskip4cm \times {{\cal P}^{(0)}_\varphi (\tilde k) }
 {{\cal P}^{(0)}_\varphi (|{\bf k}- {\bf \tilde k}|) }
 \label{finalqso}
 \,,
\eea
where in the second equality we use $d^3 {\bf \tilde k}\,=\,2 \pi \,{\tilde k}^2 \,d \tilde k\,d \mu$, as well as  properties of the combinations ${\texttt e}^{(\lambda)}({\bf k}, \tilde{\bf k})$ defined in Eq.~\eqref{intrsce} (we note that $\mu\,\equiv\,{({\bf k}\cdot {\tilde
{\bf k}})}/{(  k\,\tilde k)}
$ is the cosine of the angle between  ${\bf k}$ and ${\bf \tilde k}$). The  overall structure of the previous expression is similar to what is found in the literature  for the case of GWs induced by second order adiabatic fluctuations.
 
\smallskip
To proceed, we plug the result \eqref{finalqso} into the expression
in  Eq. \eqref{eq_ts2s}. We
use the variables $x_\star= k \tau_\star$ and $y=\tau/\tau_\star$ introduced in
Eq. \eqref{def_ys}. We also change variable from $\tilde k$ to
\be
v\,\equiv \frac{\tilde k}{ k}\,.
\ee
Moreover, we use the  following 
quantity $u$ as abbreviation in next formulas (but we do not  change variables in terms of it):
\bea
u\,\equiv\,\frac{| {\bf k}-{\bf \tilde k}|}{k}\,=\,\sqrt{1+v^2-2 \mu v}\,,
\eea
In terms of these quantitites, we find the following expression for the tensor spectrum
\bea
{\cal P}_h\left(\tau = y \tau_\star, k =\frac{ x_\star}{k_\star}\right)&=&\frac{4 \,M^4\,x_\star^2}{y^2}
\int_0^{\infty}\,d v\,\int_{-1}^{1}\,d \mu
\,{\cal P}_\varphi^{(0)}(u\,k)\,{\cal P}_\varphi^{(0)}(v\,k)\,
\frac{{\left(1-\mu^2 \right)^2}\,v^3}{u^3} \,
{\cal I}_T(y,\,x_\star)\,,
\nonumber
\\
\label{totintev3}
\eea
with
\bea
{\cal I}_T(y,\,x_\star)&\equiv&
\Big| 
\sin{(x_\star y)}
\int_{0}^{y/k_\star}\,y_1\,d y_1
\,\cos{\left(x_\star y_1 \right)}
\,\beta(y_1,x_\star, u, v)
\nonumber
\\
&&
- \cos{(x_\star y)}
\int_{0}^{y/k_\star}\,y_1\,d y_1
\,\sin{\left(x_\star y_1 \right)}
\,\beta(y_1,x_\star, u, v)
\Big|^2
\label{timein1}
\,.
\eea
In terms of the variables $x_\star,\,y$ the function $\beta$ of Eq. \eqref{defff} reads
\bea
\label{defff3}
\beta(y_1, x_\star, u, v )&=&
T(y_1, u\,x_\star)  T(y_1, v\,x_\star)-\frac{y_1^2\,\left[ \partial_{y_1} T(y_1, u\,x_\star) \right] \,\left[ \partial_{y_1} T(y_1, v\,x_\star) \right] }{\left( u^2 x_\star^2+y_1^2 \right)\left( v^2 x_\star^2+y_1^2 \right)} 
\,.
\eea

\noindent
 The 
functions $\sin{(x_\star y)}$ and $\cos{(x_\star y)}$ in front of the integrals in Eq. \eqref{timein1} are  rapidly oscillating. 
By expanding the square in Eq. \eqref{timein1}
in the late time, large $y$ limit,
the cross term 
 averages out  and the square terms $\sin^2{(x_\star y)}$
 and $\cos^2{(x_\star y)}$ average to $1/2$. Hence  we conclude 
\bea
\bar{\cal I}_T(x_\star, u, v)&=&\frac{1}{2 }\left[
\Big| 
\int_{0}^{y/k_\star}y_1 \,d y_1 
\cos{\left(x_\star y_1 \right)}
\,\beta
\Big|^2
+\Big|
\int_{0}^{y/k_\star}y_1\,d y_1
\sin{\left(x_\star y_1 \right)}
\,\beta
\Big|^2
\right]
\nonumber\\
&\equiv&\frac12\,\bar{\cal I}_{cs}^2(x_\star, u,v)
\label{timein5}
\eea
with $\bar{\cal I}_{cs}^2$ the content of the square bracket in the previous formula,
and a bar indicates averaging over many periods of the oscillatory functions $\cos(x_\star y)$, $\sin(x_\star y)$.
We numerically checked
that the integrals in Eq. \eqref{timein5} converge for large values of $y$, and in our analysis we focus on such large-$y$ regime where the integrals become independent from the value of the integral upper limits.

\smallskip

We now use the tensor power spectrum ${\cal P}_h$ of Eq. \eqref{totintev3} to build the quantity $\Omega_{\rm GW}$, proportional to the gravitational energy density 
per log frequency interval. $\Omega_{\rm GW}$ is a standard  parameter
which controls the magnitude of SGWB in cosmological setting. It is related to the tensor spectrum by
\be
\Omega_{\rm GW}\,=\,\frac{1}{24}\,\frac{k^2}{a^2 H^2}\,{\cal P}_h\,.
\ee
During RD, 
$a H=1/\tau$. We collect the previous formulas, and obtain
\bea
\label{intOGW}
\Omega_{\rm GW}(k)
&=&
\frac{M^4}{12}\,x_\star^4\,
\int_0^{\infty}\,d v\,\int_{-1}^{1}\,d \mu
 \,
\bar{\cal I}_{cs}^2(x_\star, u, v)
\,{\cal P}_\varphi^{(0)}(u\,k)\,{\cal P}_\varphi^{(0)}(v\,k)\,
\frac{{\left(1-\mu^2 \right)^2}\,v^3}{u^3}\,,
\eea
where we recall
$
u\,=\,\sqrt{1+v^2-2 \mu v}
$.  In view
of a numerical evaluation of the integrals, it is convenient to perform another change of variables, following \cite{Kohri:2018awv}:
\bea
t&\equiv&u+v-1
\hskip0.5cm,\hskip0.5cm
s\,\equiv\,u-v\,.
\eea
\begin{figure}[t!]
    \centering
    \includegraphics[width=0.58\linewidth]{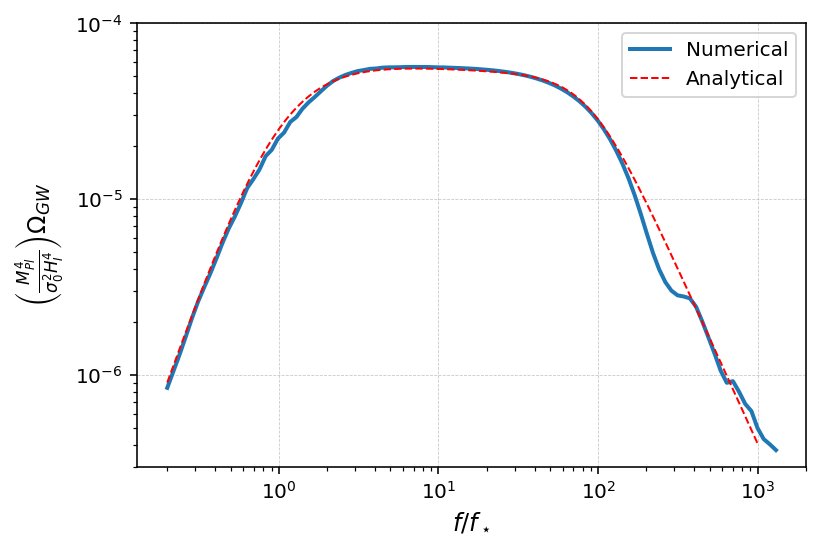}
    \caption{ \small 
Representation of the GW energy density $\Omega_{\rm GW}$ of Eq. \eqref{intOGW3},   divided by $\sigma_0^2\,H^4_I/M_{\rm Pl}^4$ (continuous blue line).  We include also
the analytical fitting function \eqref{eq_OGWfit} (dashed red line).}
    \label{fig:OGW}
\end{figure}
The GW energy density becomes
\bea
\Omega_{\rm GW}(k)
&=&
\frac{M^4\,x_\star^4}{12}\,
\int_0^{\infty}\,d t\,\int_{0}^{1}\,d s
\,
\left[\frac{t (2+t) (s^2-1)}{(1-s+t) (1+s+t)}
\right]^2
 \,\bar{\cal I}_{cs}^2(x_\star, u, v)\,
\,{\cal P}_\varphi^{(0)}(u\,k)\,{\cal P}_\varphi^{(0)}(v\,k)\,
\nonumber
\\
\label{intOGW2}
\eea
with
$
u\,=\,(t+s+1)/2$,  
$v\,=\,(t-s+1)/2$. 
The integral \eqref{intOGW2}
resembles  formulas for curvature scalar
induced GW -- only the structure
of the kernel $\bar{\cal I}_{cs}^2$ is different, compare e.g. with  \cite{Kohri:2018awv}.

\smallskip
So far, we have not specified the amplitude of the sourcing fields. We can now make use of the results of Section \ref{sec_rev} for 
the primordial spectrum ${\cal P}_\varphi^{(0)}$ of the longitudinal scalar mode. Such spectrum
is scale invariant, and depends on the inflationary Hubble parameter $ H_I$ and vector mass $M$ \eqref{pphiin2}: 
\be
\label{spec_ls}
{\cal P}_\varphi^{(0)}(k)\,=\,
{\sigma}_0\,\frac{ H_I^2}{4 \pi^2 M^2}
\,.
\ee
We include  an overall, model-dependent  factor $\sigma_0$ in the previous formula.
As discussed around Eq. \eqref{pphiin2}, it aims to take into account of possible refinements of the original
model \cite{Graham:2015rva} associated with phase transitions,  non-standard evolution history or reheating after inflation, or more exotic inflationary evolution including  non-minimal
couplings with gravity. We do not commit  on
explicit scenarios, and we are
going to make use of $\sigma_0$ as useful tunable  parameter in our  set-up.~\footnote{A particular example able to generate a large $\sigma_0$ is discussed after Eq. \eqref{vedb}.}

The GW energy density results (reinstating the  Planck mass, and expressing it in terms of frequency since $x_\star=k/k_\star=f/f_\star$, with $k\,=\,2 \pi\, f $) 
\bea
\Omega_{\rm GW}(f)
&=&\sigma_0^2\,\left( \frac{H_I}{M_{\rm Pl}}\right)^4
\left[
\frac{f^4}{192\,\pi^4\,f_\star^4}\,
\int_0^{\infty}\,d t\,\int_{0}^{1}\,d s
\,
\left(\frac{t (2+t) (s^2-1)}{(1-s+t) (1+s+t)}
\right)^2
 \,\bar{\cal I}_{cs}^2({f}/{f_\star}, u, v)\right]
\nonumber
\\
\label{intOGW3}
\eea
We numerically evaluate the model-independent, dimensionless quantity inside square brackets, and we represent it in Fig \ref{fig:OGW} as function of $f/f_\star$ -- the fiducial frequency being $2 \pi\,f_\star\,=\, a_\star\, M$ (see
Eq. \eqref{def_kst}). Notice that in representing $\Omega_{\rm GW}$ we factorise powers of   the parameters $\sigma_0$ and of  the value of the Hubble parameter during inflation.

\smallskip

The resulting frequency profile plotted in  Fig \ref{fig:OGW}  is quite broad, reflecting
the scale invariant primordial spectrum for the longitudinal scalar $\varphi$ as given
in Eq. \eqref{spec_ls}. We find a good analytical fit in terms of a double broken
power law as
\begin{equation}
\label{eq_OGWfit}
    \left(\frac{M^4_{Pl}}{\sigma_0^2\,H^4_I}\right)\times \Omega_{\rm GW}\left(x_\star= {f}/{f_\star}\right)\,=\,6.25\times 10^{-5}\,\,\frac{x_\star^{2.6}}{\left(1+x_\star^{2} \right)^{1.325}}\left(1+ \frac{x_\star^3}{7.29\times 10^5}  \right)^{-0.65}
\end{equation}
The maximal amplitude of the combination in the right hand side of Eq. \eqref{eq_OGWfit}
is of order $5 \times 10^{-5}$, and its profile remains almost constant at such maximal values in the range $3\,< \,f/f_\star\,<\,60$. The profile of 
$\Omega_{\rm GW}(x_\star)$ grows as $f^{2.6}$ in the infrared,
and it decreases as $f^{-2}$ in the ultraviolet.

Depending on the values of $\sigma_0$, $H_I$, and $f_\star$ our results
 have different phenomenological consequences:  which we explore in the next section.

\section{Phenomenological consequences}
\label{sec_pheno}

We now collect the results of the previous sections and discuss their phenomenological consequences for the physics of the induced GW from longitudinal vector DM. The longitudinal vector mode $A_L$ has a spectrum
peaked around the scale $k_\star = a_\star M$, with $M$ the vector mass during RD, while $a_\star$ is the value of the scale factor which corresponds to $H(a_\star)=M$  -- see Section \ref{sec_rev}. 
We setting the present value of the scale factor to one, $a_0=1$, so that the momentum scale $k_\star$ is mapped to the frequency $f_*$ via
\bea
2 \pi\,f_\star&=& k_\star = {M\,a_\star}
\,=\,\frac{M}{z_{\rm eq}}\,
\left(\frac{H_{\rm eq}}{H_*} \right)^{1/2}
\,=\,\frac{
\sqrt{M\,H_{\rm eq}}
}{z_{\rm eq}}
\,=\,4.95\,\times\,10^{-18}\,{\rm eV}
\,\left( \frac{M}{\rm eV}\right)^{1/2}
\,,
\eea
where we make use of $H_{\rm eq}= 3\times 10^{-28}$
 eV and $z_{\rm eq}=3400$ \cite{Planck:2018vyg}. ``Converting'' eV to Hertz, we get
\be
f_\star\,=\,1.2
\,\left( \frac{M}{10^6\,{\rm eV}}\right)^{1/2}\,{\rm Hz}
\label{eq_ffr}
\,.
\ee
Actually,
as we observe from Fig \ref{fig:OGW}, the induced GW spectrum grows and is amplified at frequencies slightly larger than $f_\star$, acquiring its maximal values for frequencies in the order $c_0
f_\star$ with $c_0 
\simeq {\cal O}(10)$. 
 To proceed, we 
 recall formula \eqref{eq_ratioen} which provides the ratio of longitudinal vector $A_L$ versus the dark matter
energy density today
\bea
\frac{\rho_A}{\rho_{\rm DM}}
&=&12.7\,\sigma_0\,\left(\frac{\rm eV}{H_{\rm eq}}
\right)^{\frac12}
\,\left(\frac{M}{10^6\,{\rm eV}}
\right)^{\frac12}
\,\left(
\frac{H_I}{M_{\rm Pl}}
\right)^2 
\nonumber
\\
&=&5.7\,\times 10^{14}\,\sigma_0\,
\left(\frac{M}{10^6\,{\rm eV}}
\right)^{\frac12}
\,\left(
\frac{H_I}{M_{\rm Pl}}
\right)^2\,,
\label{eq_fdm}
\eea
where $H_I$ is the Hubble parameter
during inflation, and $\sigma_0$ is the model-dependent
overall factor in the longitudinal scalar 
spectrum, see Eq. \eqref{pphiin2} and related discussion. Finally, the result summarized by  Figure  \ref{fig:OGW} indicates that the induced GW signal can obtain the maximum value 
\bea
\label{eq_fog}
\Omega_{\rm GW}^{\rm max}
&\simeq&5\times10^{-5}\,{\sigma}_0^2
\,\left(\frac{H_I}{M_{\rm Pl}} \right)^4 \;. 
\eea

The last three equations can be employed to study the phenomenology of the model for different values of the model parameters. Let us first focus on the original model ~\cite{Graham:2015rva}, for which $\sigma_0 = 1$ and we demand $\rho_A/\rho_{\rm DM}=1$. Current CMB polarisation experiments
set the bound $r \le 0.03$ on the tensor-to-scalr ratio at large scales \cite{BICEP2:2018kqh}, and, correspondingly, on the scale of inflation $H_I/M_{\rm Pl}\le 10^{-5} \sqrt{r/0.03}$.  Let us choose a reference value of $H_I/M_{\rm Pl}=10^{-6}$. Then, Eq.~\eqref{eq_fdm} indicates that the longitudinal vector coincides with the dark matter if its mass is given by $M\,\simeq\,3$ eV. From Eq.~\eqref{eq_ffr} we then see that this mass corresponds to the characteristic GW frequency $f_\star \simeq 2\times 10^{-3}$ Hz. However, the resulting $\Omega_{\rm GW}^{\rm max}
 \,=\,10^{-29}$, too small to be detected with a milli-Hertz interferometer as LISA \cite{LISA:2024hlh}. To obtain an observable signal, we need to tune the parameter $\sigma_0$ to larger values.

\smallskip

Combining eqs \eqref{eq_ffr}, \eqref{eq_fdm}, and \eqref{eq_fog}, we find the consistency condition 
\bea
\label{eq_const}
\frac{\Omega^{\rm max}_{\rm GW}}{10^{-10}}
&=&
\left(
\frac{c_0}{67}\right)^2
\left(
\frac{\rho_A}{\rho_{\rm DM}}
\right)^2
\,\left(\frac{10^{-10}\,{\rm Hz}}{c_0\,f_{\star}}
\right)^2\,,
\eea
where we include $c_0$ as an overall factor in front of
the frequency $f_\star$, in order to take into account that the peak $\Omega_{\rm GW}^{\rm max}$  is shifted with respect to $f_\star$: see Fig \ref{fig:OGW}
and the discussions after Eq. \eqref{eq_OGWfit} and Eq. \eqref{eq_ffr}. Interestingly, the  relation
\eqref{eq_const} is  {\it independent}
of the model parameters $M$, $H_I$, and $\sigma_0$.  It only relies on the hypothesis
that the longitudinal scalar spectrum ${\cal P}_\varphi$ is scale invariant during inflation,
as in Eq. \eqref{pphiin2}. Eq. \eqref{eq_const} allows us to identify the most relevant parameter regions, and, in particular, it leads to the conclusion that models that provide a signalled peaked comparatively smaller frequencies $f_*$ have the chance to produce a comparatively larger GW signal, provided of course $\sigma_0$ is large enough. For definiteness, let us consider
three representative examples:
\begin{figure}[t!]
    \centering
    \includegraphics[width=0.3\linewidth]{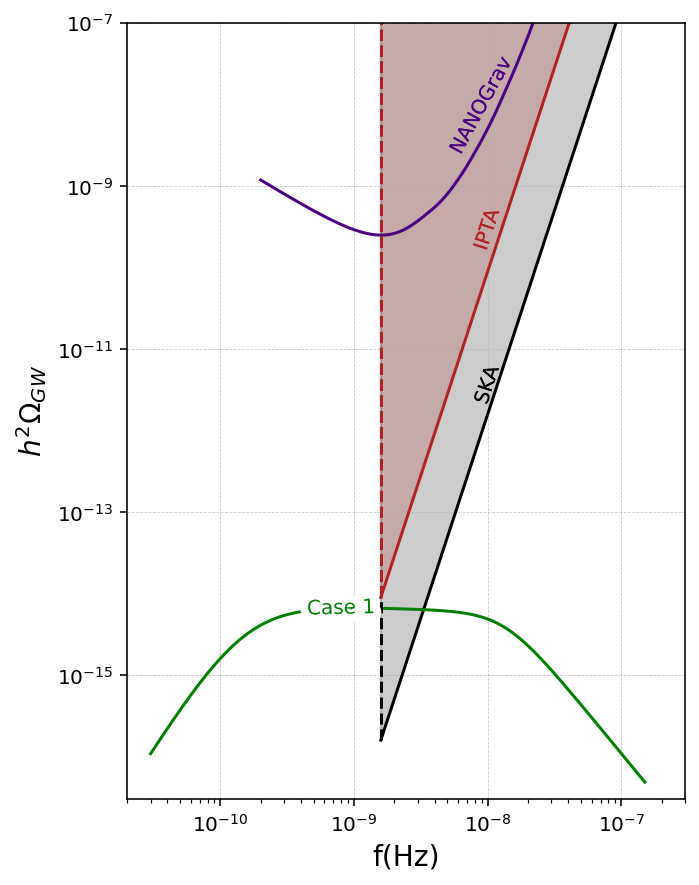}
    \includegraphics[width=0.3\linewidth]{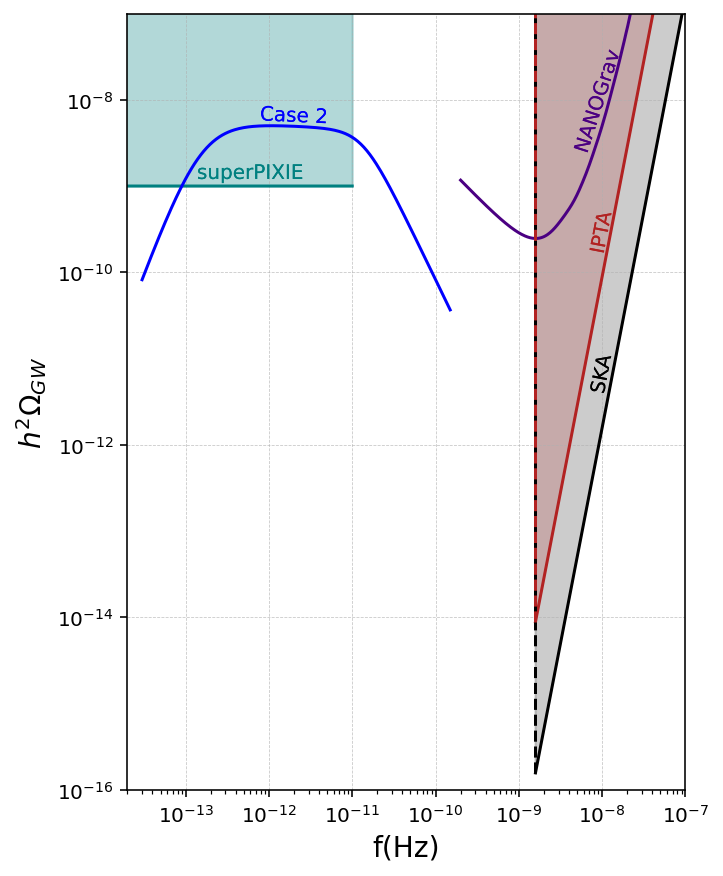}
        \includegraphics[width=0.3\linewidth]{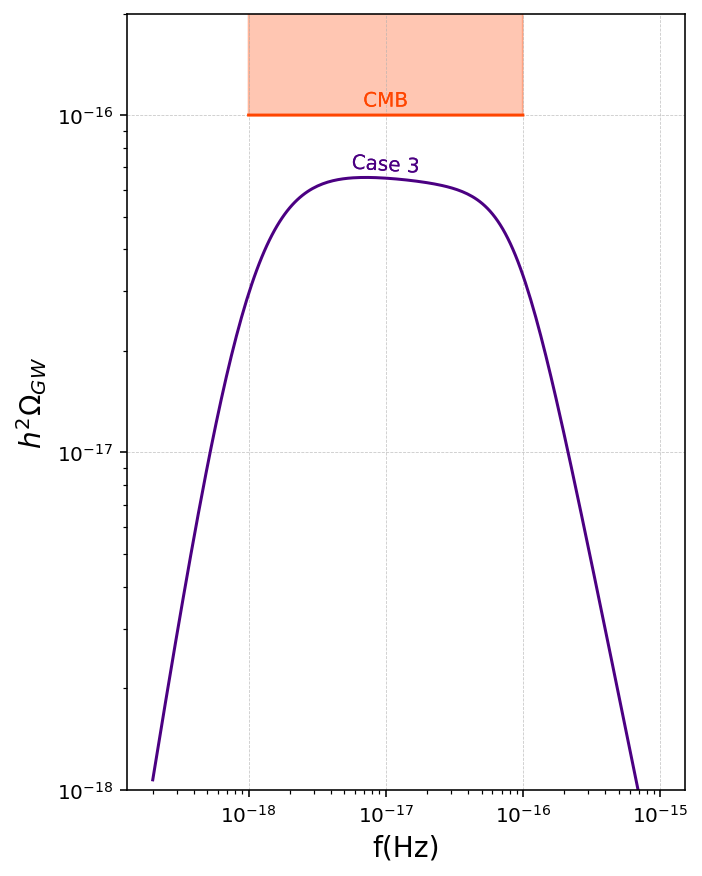}
    \caption{\small Representation of the three cases discussed in the main text. {\bf Left panel}: Case 1. We take the sensitivity curves
    for nanoHz GW experiments from \cite{Schmitz:2020syl}. {\bf Middle panel}: Case 2. Sensitivities curves associated to redshift space-distortion experiments 
    are taken from \cite{Kite:2020uix}.  {\bf Right panel}: Case 3.  
    }
    \label{fig:case1}
\end{figure}
\begin{itemize}
\item[-] {\bf Case 1: Peak at nano-Hertz scales.} We choose $f_\star=1.5 \times 10^{-10} \, {\rm Hz}$ and we demand that the longitudinal vector forms all of the dark matter. Then Eq. \eqref{eq_const} results in the amplitude $\Omega^{\rm max}_{\rm GW}=10^{-14}$. In turn, Eq.~\eqref{eq_ffr} indicates the vector mass $M \simeq 1.6 \times 10^{-14} eV$, while Eq.~\eqref{eq_fog} shows that the parameter $\sigma_0$ needs to satisfy $\sigma_0 \simeq 1.4 \times 10^7 \left( \frac{10^{-6} \, M_p}{H_I} \right)^2$. See Fig \ref{fig:case1}, left panel, where we compare the analytical fit of $\Omega_{\rm GW}$
in Eq. \eqref{eq_OGWfit}  with  nominal sensitivity curves of (current and future) PTA experiments, taken from \cite{Schmitz:2020syl}. In the future, SKA should be able to probe the nearly flat, central region of the $\Omega_{\rm GW}$ profile, provided that  it can resolve this 
specific cosmological backgrounds against
an astrophysical signal. Possible methods might rely on the properties of SGWB anisotropies  \cite{LISACosmologyWorkingGroup:2022kbp,Cusin:2022cbb,Baker:2019ync}.
Further measurements can also benefit from  astrometry  observations (see  \cite{Book:2010pf} for a review),
and by the synergy of astrometry and PTA \cite{Cruz:2024diu}. 
\item[-] {\bf Case 2: Peak at intermediate scales.}
Let us instead consider a peak frequency $f_\star\,=\, 1.5\times 10^{-13}$  Hz, at intermediate scales between CMB and PTA experiments. Assuming again that the longitudinal vector forms all of the dark matter, then Eq. \eqref{eq_const} results in the amplitude $\Omega^{\rm max}_{\rm GW}=10^{-8}$. Eq.~\eqref{eq_ffr} points now to the vector mass $M=1.5 \times 10^{-20}$ eVm which makes the light vector a `fuzzy dark matter' candidate, see e.g. \cite{Ferreira:2020fam} for a review. From Eq.~\eqref{eq_fog}, we must require $\sigma_0 \simeq 1.4 \times 10^{10} \left( \frac{10^{-6} \, M_p}{H_I} \right)^2$. The central panel of Fig. \ref{fig:case1} shows a representation of the analytical fit of $\Omega_{\rm GW}$ in Eq. \eqref{eq_OGWfit} against the perspective sensitivity of super PIXIE -- taken from \cite{Kite:2020uix} -- an evolution of the PIXIE experiment \cite{Kogut:2011xw,Kogut:2019vqh,Chluba:2019nxa} to measure redshift space-distortions. This result demonstrates the importance of finding methods to measure  GW at intermediate frequency ranges between CMB and PTA experiments. In fact, redshift space-distortions is a promising probe for testing the SGWB induced by  models of longitudinal vector DM.
\item[-] {\bf Case 3: Peak at CMB scales.} Present bounds on
the amount of primordial 
B-modes in CMB polarisation experiments  set constraints
on regions of parameter space  enhancing
the GW spectrum at small frequencies. For example, 
 consider scenarios potentially able to amplify
the SGWB magnitude  at CMB scales  $f_{\rm CMB}\simeq 10^{-18}$ Hz --
corresponding to tiny vector masses $M \simeq 7 \times 10^{-31}$ eV, see Eq. \eqref{eq_ffr}. On such frequency ranges, current B-mode constraints impose $\Omega_{\rm GW}(f_{\rm CMB})
\,\le\,(r/0.01)\,\times 10^{-15}$  \cite{BICEP2:2018kqh}. Choosing for instance $r=10^{-3}$ results in $\Omega_{\rm GW}^{\rm max}=10^{-16}$, which is compatible with only a very small vector field energy density, ${\rho_A}/{\rho_{\rm DM}}\,\le\, 7\times 10^{-10}$, see Eq.~\eqref{eq_const}. Future experiments as LiteBIRD can gain further orders of magnitude in sensitivity, see e.g. 
 \cite{Campeti:2020xwn}. It is remarkable that GW experiments are able to set so stringent constraints
 on tiny amounts of longitudinal vector in the DM budget.
\end{itemize}

What is particularly interesting is 
that the GW mechanism of production that we analysed is {independent} of any possible coupling of the vector
$A_\mu$ to Standard Model physics. It  holds even if there are no couplings 
at all with the Standard Model (besides gravity) -- or very tiny ones \cite{Gherghetta:2019coi}. 
It is interesting that the induced GW spectrum can be enhanced at frequencies below 
the nano-Hertz, see Eq. \eqref{eq_const}, setting new targets to GW experiments. Nevertheless, the requirement to get detectable signals (say values of $\Omega_{\rm GW}$ in the $10^{-10}$ regime) requires us to choose large values of $\sigma_0$. In principle this is possible by elaborating extensions of the original scenario \cite{Graham:2015rva} as discussed in the paragraphs
around Eq. \eqref{pphiin2}. Possibly, the simplest way to do so is by considering scenarios in which 
the vector mass deep in the inflationary era, $m_{\rm inf}$ is much smaller  from the one during radiation domination, call it $M$.
This condition changes the couplings to gravitational waves and lead to an effective value of $\sigma_0=(M/m_{\rm inf})^2$, as commented after Eq. \eqref{vedb}.  Possible mechanisms related with phase transitions are discussed in \cite{Salehian:2020asa}, but more model building efforts will needed to better explore these possibilities and place them in a firmer theoretical setting. 
  
Notice also that our formulas indicate that a sufficiently large GW signal is associated with very tiny masses for the vector boson,  in regimes usually considered as fuzzy dark matter. In this case, there are other complementary gravitational probes of such instances, associated with phenomena like black hole
superradiance from light vectors \cite{Baryakhtar:2017ngi,Siemonsen:2019ebd}. At the astrophysics and gravitational level, it is also worth noticing that light vector
bosons, like the ones considered, can form coherent fuzzy dark matter structures, and compact objects related to Proca stars \cite{Brito:2015pxa,SalazarLandea:2016bys,Sanchis-Gual:2017bhw,Tasinato:2022vop,Atkins:2023axs,Tasinato:2014eka,Chagoya:2016aar,Chagoya:2017fyl}. 
It will be interesting to further investigate realistic particle physics constructions behind light longitudinal
vector dark matter models, and understand at what extent improved GW detectors  can constrain (or probe!) their predictions.
\section{Outlook}
\label{sec_con}
We discussed a new gravitational wave probe of longitudinal vector dark matter. We computed
the stochastic gravitational wave background produced at second
order in fluctuations in scenarios of  vector dark matter  based on   \cite{Graham:2015rva}.
Our approach is  similar in spirit to  analysis on second order GW produced by primordial black hole models -- this time
applied to a particle physics  dark matter set-up, and to the case
of non-adiabatic fluctuations. 
We have shown that the GW spectrum is amplified at very small frequencies, at, or below, the nano-Hertz. 
Although possibly detectable at nano-Hertz scales, 
GW
from vector dark matter constitute a particularly  interesting target for very low frequency GW experiments, able to detect GW say in the region around $10^{-12}$ Hertz.  The original DM model is very minimal -- the action
has only the vector mass as free parameter -- but over the years several generalisations and refinements
have been considered. In fact, only scenarios extending  \cite{Graham:2015rva} lead to the hope of producing  
a SGWB amplitude detectable by future experiments.  Models characterised by a small vector
mass are more likely to lead to a detectable SGWB signal, and can be further tested
by other GW probes as black hole superradiance.
Hence, it would be very interesting to continue   model building efforts  in order to identify well motivated  scenarios able to lead to a measurable GW signal. Since we only focussed
on production of GW during radiation domination, 
it would
also be interesting to study GW production during matter domination, or in non-standard cosmologies. Also, since the vector action is relatively simple, it would also be interesting to study the effects of non-linearities.

We are setting challenging targets, both at the
 theoretical and  at the experimental level. However, recall that  the only known coupling of dark matter with Standard Model is through gravity: it might be that for some reasons such coupling is the only one chosen by Nature.  In this   case, further exploring the gravitational wave probe we identified -- based on gravity only and in principle testable with GW experiments --
is certainly a well motivated pursuit.

\bigskip
\subsection*{Acknowledgments}

It is a pleasure to thank Ogan \"Ozsoy for discussions on related topics. The work of AMB and GT is partially funded by STFC grant ST/X000648/1.  
M.P. acknowledges support from Istituto
Nazionale di Fisica Nucleare (INFN) through the Theoretical Astroparticle Physics (TAsP) project,
and from the MIUR Progetti di Ricerca di Rilevante Interesse Nazionale (PRIN) Bando 2022 - grant
20228RMX4A. For the purpose of open access, the authors have applied a Creative Commons Attribution licence to any Author Accepted Manuscript version arising. Research Data Access Statement: No new data were generated for this manuscript.

{\small

\providecommand{\href}[2]{#2}\begingroup\raggedright\endgroup
}

\end{document}